# The Toroidal Iron Atmosphere of a Protoneutron Star: Numerical Solution

**V. S. Imshennik*, K. V. Manukovskii, and M. S. Popov**

*Institute for Theoretical and Experimental Physics, ul. Bol'shaya Cheremushkinskaya 25, Moscow, 117259 Russia*

**Abstract**—A numerical method presented by Imshennik *et al.* (2002) is used to solve the two-dimensional axisymmetric hydrodynamic problem on the formation of a toroidal atmosphere during the collapse of an iron stellar core and outer stellar layers. An evolutionary model from Boyes *et al.* (1999) with a total mass of $25M_\odot$ is used as the initial data for the distribution of thermodynamic quantities in the outer shells of a high-mass star. Our computational region includes the outer part of the iron core (without its central part with a mass of $1M_\odot$ that forms the embryo of a protoneutron star at the preceding stage of the collapse) and the silicon and carbon–oxygen shells with a total mass of $(1.8–2.5)M_\odot$. We analyze in detail the results of three calculations in which the difference mesh and the location of the inner boundary of the computational region are varied. In the initial data, we roughly specify an angular velocity distribution that is actually justified by the final result—the formation of a hydrostatic equilibrium toroidal atmosphere with reasonable total mass, $M^{\rm tot} = (0.117–0.122)M_\odot$, and total angular momentum, $J^{\rm tot} = (0.445–0.472) \times 10^{50}$ erg s, for the two main calculations. We compare the numerical solution with our previous analytical solution in the form of toroidal atmospheres (Imshennik and Manukovskii 2000). This comparison indicates that they are identical if we take into account the more general and complex equation of state with a nonzero temperature and self-gravitation effects in the atmosphere. Our numerical calculations, first, prove the stability of toroidal atmospheres on characteristic hydrodynamic time scales and, second, show the possibility of sporadic fragmentation of these atmospheres even after a hydrodynamic equilibrium is established. The calculations were carried out under the assumption of equatorial symmetry of the problem and up to relatively long time scales (∼10 s).

## INTRODUCTION

The theory of the spherically symmetric gravitational collapse of iron stellar cores is known to have led to a clear separation of the collapse into two stages (see, e.g., Imshennik and Nadyozhin 1982). At the first stage, the inner part of the iron core with a mass $M_{i\rm Fe} \leq 1M_\odot$ (Imshennik 1992) or, more precisely, with a mass of $(0.6–0.8)M_\odot$ (Nadyozhin 1998) collapses homologically. The remaining outer part of the iron core with a mass $M_{e\rm Fe} = M_{\rm Fe} - M_{i\rm Fe} \cong (0.4 - 1.4)M_\odot$ in the known range of iron core masses $M_{\rm Fe} = (1.2 - 2.0)M_\odot$ lags well behind its inner part in contraction. In a rough approximation, it may even be assumed to maintain a hydrostatic equilibrium of the initial state. This is especially true for the outer layers of a high-mass star. Imshennik *et al.* (2002) reduced the question about the hydrodynamic behavior of the outer part of the iron core (the second stage of its collapse) and other outer layers of a collapsing star to the problem of their accretion from a hydrostatic equilibrium onto the embryo of a protoneutron star (PNS) with a mass of about $1M_\odot$. The end of the first stage of iron core collapse was taken as the initial time of hydrodynamic post-shock accretion; the corresponding problem was formulated by Brown *et al.* (1992). This simplification of the problem allowed us, first, to exclude the complex neutrino processes from consideration (i.e., to restrict our analysis to the approximation of ideal hydrodynamics), and, second, to include the hydrodynamic processes in the remaining onion stellar structure that surrounds the iron core of a high-mass star with a total mass

*E-mail: imshennik@vxitep.itep.ru

$M_{\rm MS} \geq 10M_{\odot}$.[1] In this paper, we continue to numerically analyze this post-shock accretion, but now we take into account the rotation effects by solving the two-dimensional axisymmetric problem.

The main result of the gravitational collapse of a rapidly rotating iron stellar core can be a binary system of neutron stars (Imshennik 1992) and its surrounding iron gaseous medium that did not enter into the composition of the rotating PNS at the second stage of the collapse. In this case, the rotating PNS is generally unstable against fragmentation and turns into a binary system of neutron stars (Aksenov *et al.* 1995). Based on a quasi-one-dimensional model, Imshennik and Nadyozhin (1992) showed that, with a sufficiently rapid initial rotation of an $M_{\rm Fe} = 2M_{\odot}$ iron core and a full allowance for the neutrino processes, an iron atmosphere with a mass of $(0.1{-}0.2)M_{\odot}$ distributed almost uniformly throughout the iron core with an initial outer radius $R_{\rm Fe} = 4.38 \times 10^8$ cm, i.e., with a mean density $\rho_{e\rm Fe} \cong (5.66 \times 10^5{-}1.13 \times 10^6)$ g cm$^{-3}$, is formed. The presumed long existence of this binary system of neutron stars, possibly for several hours (Imshennik and Popov 1994), made the problem of steady iron gas accretion onto this binary (actually, onto its more massive component located almost at the stellar center) of relevant interest. For the one-dimensional accretion of a cold (with zero temperature) iron gas with quasi-one-dimensional rotation, Imshennik and Popov (2001) managed to generalize the standard steady-state solution by Bondi (1952) due to the very simple form of the specific enthalpy for the degenerate electron gas of an iron atmosphere with the complete ionization of iron atoms and arbitrary electron relativity. Application of this solution to the physical conditions near PNSs (with masses $(1.4{-}1.8)M_{\odot}$) showed that the accretion rate was very high, $(20{-}30)M_{\odot}$ s$^{-1}$ at the above typical densities $\rho_{e\rm Fe}$ and a neutron star mass of $1.8M_{\odot}$. On the other hand, the steady-state accretion solution obtained does not fit into the iron core; besides, the time it takes for this solution to be established is too long. In short, it is not applicable to our problem on the second stage of the collapse of the outer iron core (Imshennik and Popov 2001). The quasi-one-dimensional rotation effects taken into account in the above paper proved to be negligible because of the severe constraints on the specific angular momentum $j_{\infty}$ (at infinity!) specified in addition to the density $\rho_{\infty}$. At the same time, based on ideal two-dimensional hydrodynamics, we were able to analytically construct a steady-state solution with the same equation of state for a cold iron gas in the form of toroidal atmospheres with a given arbitrary rotation law (Imshennik and Manukovskii 2000).

A natural continuation of the above studies of analytical solutions seems to be the application of our numerical method of solution (Imshennik *et al.* 2002) to the problem of postshock accretion near neutron stars, using a complete equation of state with temperature effects (arbitrary electron degeneracy, the appearance of positrons, the presence of a nuclide gas and blackbody radiation), and a consistent allowance for the gravitational interaction, including the two-dimensional self-gravitation of the accreted matter. As was done previously (Imshennik *et al.* 2002), the hydrostatic equilibrium solutions obtained in evolutionary calculations (Boyes *et al.* 1999) and modified to incorporate the quasi-one-dimensional rotation effects can also be used as the initial time and the initial conditions that describe the statement of the postshock accretion problem (see above). Note that the accuracy of the hydrostatic equilibrium of these outer layers, including the action of centrifugal forces is not of such fundamental importance as it was in Imshennik *et al.* (2002), because it essentially does not matter precisely what time is taken as the initial one in the unsteady-state hydrodynamic simulation of the outer layers of a high-mass star under consideration.

As we show in this paper, the hydrodynamic postshock accretion actually gives rise to toroidal structures at the locations of the outer layers of an iron stellar core, in agreement with the analytical solutions from Imshennik and Manukovskii (2000). Thus, we clearly prove the hydrodynamic stability of such toroidal iron atmospheres against arbitrarily small two-dimensional perturbations. As regards the steady one-dimensional accretion, it does not take place during the postshock accretion: instead, significant accretion onto the embryo of a PNS (with an initial mass of approximately $1M_{\odot}$) takes place in the near-axis region for part of the matter with the sufficiently low specific angular momenta specified in the initial conditions (with a total mass of about $0.8M_{\odot}$).

[1] Imshennik *et al.* (2002) showed that the previously assumed influence of the decrease in the gravitational mass of a PNS due to intense neutrino energy losses during collapse was negligible: it does not lead to the emersion of the outer stellar layers in a wide range of total masses, $(12{-}25)M_{\odot}$. Unfortunately, another effect, the imitation of nucleon bubbles embedded in an initial spherically symmetric configuration under equilibrium conditions, was considered with the erroneous overestimation of their total energy attributable to the unacceptably abrupt changes in initial specific internal energy at the outer boundary of the nucleon bubbles. After the rectification of these unfortunate inaccuracies (a several-fold decrease in initial specific internal energy), the emersion effect of the outer layers also vanished, and it was generally interpreted as a small correction to the adopted equation of state because of the energy release of free-nucleon recombination into iron.

This accretion is not only non-one-dimensional but also unsteady, with the ejection of whole fragments of matter from the toroidal atmosphere, a curious effect that was found in these calculations.

## FORMULATION OF THE PROBLEM

### *The System of Equations, the Equation of State, and the Numerical Method of Solution*

In most cases, the hydrodynamic behavior of the envelope of a high-mass star can be described by the system of equations of ideal hydrodynamics. In our axisymmetric case $\left(\frac{\partial}{\partial \varphi}, g_\varphi = 0\right)$, this system in spherical coordinates $(r, \theta, \varphi)$ is

$$\frac{\partial \rho}{\partial t} + \frac{1}{r^2}\frac{\partial}{\partial r}(r^2 \rho V_r) + \frac{1}{r \sin\theta}\frac{\partial}{\partial}(\sin\theta \rho V_\theta) = 0, \quad (1)$$

$$\frac{\partial \rho V_r}{\partial t} + \frac{1}{r^2}\frac{\partial}{\partial r}(r^2 \rho V_r^2) + \frac{1}{r \sin\theta}\frac{\partial}{\partial \theta}(\sin\theta \rho V_r V_\theta) \quad (2)$$

$$+ \frac{\partial P}{\partial r} - \frac{\rho(V_\theta^2 + V_\varphi^2)}{r} = \rho g_r,$$

$$\frac{\partial \rho V_\theta}{\partial t} + \frac{1}{r^2}\frac{\partial}{\partial r}(r^2 \rho V_r V_\theta) + \frac{1}{r \sin\theta}\frac{\partial}{\partial \theta}(\sin\theta \rho V_\theta^2) \quad (3)$$

$$+ \frac{1}{r}\frac{\partial P}{\partial \theta} + \frac{\rho V_r V_\theta}{r} - \frac{\rho V_\varphi^2 \cot\theta}{r} = \rho g_\theta,$$

$$\frac{\partial \rho V_\varphi}{\partial t} + \frac{1}{r^2}\frac{\partial}{\partial r}(r^2 \rho V_r V_\varphi) + \frac{1}{r \sin\theta}\frac{\partial}{\partial \theta}(\sin\theta \rho V_\theta V_\varphi) \quad (4)$$

$$+ \frac{\rho V_\varphi}{r}(V_r + \cot\theta V_\theta) = 0,$$

$$\frac{\partial \rho E}{\partial t} + \frac{1}{r^2}\frac{\partial}{\partial r}(r^2 V_r(\rho E + P)) \quad (5)$$

$$+ \frac{1}{r \sin\theta}\frac{\partial}{\partial \theta}(\sin\theta V_\theta(\rho E + P)) = \rho(V_r g_r + V_\theta g_\theta),$$

where $E = \varepsilon + \frac{\mathbf{v}^2}{2}$ is the sum of the specific internal ($\varepsilon$) and kinetic energies. The acceleration of gravity is $\mathbf{g} = \mathbf{g}_{\rm ns} + \mathbf{g}_{\rm atm}$; i.e., it is the sum of two components. The first component is attributable to the gravitational field that is produced by the PNS located exactly at the coordinate origin:

$$\mathbf{g}_{\rm ns} = \left(-\frac{GM_{\rm ns}}{r^2}, 0, 0\right). \quad (6)$$

The second component is attributable to the intrinsic gravitational field of the iron atmosphere. The acceleration of this field is defined by the standard equation $\mathbf{g}_{\rm atm} = -\nabla\Phi$, and the potential satisfies the Poisson equation

$$\Delta\Phi = 4\pi G\rho. \quad (7)$$

An efficient algorithm convenient for use in the finite-difference scheme of integrating the hydrodynamic equations on a stationary mesh in spherical coordinates (Aksenov 1999) is used to solve the Poisson equation (7) and determine the gravitational acceleration $\mathbf{g}_{\rm atm}$. Thus, in the problem under consideration, we abandon the Roche approximation that was used previously for simplicity (Imshennik and Manukovskii 2000; Imshennik and Popov 2001) and take into account the atmospheric self-gravitation effects. Note that, in general, they are of minor importance.

In this paper, as in our previous paper (Imshennik *et al.* 2002), we use the equation of state for the matter treated as a mixture of a perfect Boltzmann gas of nuclei with a perfect Fermi–Dirac electron–positron gas and blackbody radiation. In a wide temperature range, the matter is assumed to be a mixture of a baryonic component that includes free nucleons ($n$, $p$) and helium ($^4_2$He) and iron ($^{56}_{26}$Fe) nuclides, and a Fermi–Dirac electron–positron gas with blackbody radiation. The equation of state obeys the nuclear statistical equilibrium (NSE) conditions with a constant ratio of the mass fractions of neutrons and protons, including those bound in the helium and iron nuclides, $\theta_0 = 30/26$ (Imshennik *et al.* 2002). As previously, we use a tabulated equation of state and precompute the thermodynamic functions with a high accuracy. This method allows complex calculations at each step of the solution of system (1)–(5) to be avoided.

The numerical solution of our problem is based on the popular and universal PPM method. This method uses the Eulerian finite-difference scheme (Colella and Woodward 1984) and is a modification of Godunov's method (Godunov *et al.* 1976). The salient features of the method used were given previously (Imshennik *et al.* 2002). In particular, the real equation of state in the numerical method is locally simulated by the so-called binomial equation of state

$$P = [(\bar\gamma - 1)\varepsilon + c_0^2]\rho - \rho_0 c_0^2; \quad (8)$$

the constants $\bar\gamma$, $c_0^2$, and $\rho_0$ are determined by the pressure and its derivative at a given point $(\rho, \varepsilon)$:

$$c_0^2 = \left(\frac{\partial P}{\partial \rho}\right)_\varepsilon - \frac{\varepsilon}{\rho}\left(\frac{\partial P}{\partial \varepsilon}\right)_\rho, \quad (9)$$

$$\rho_0 c_0^2 = \rho\left(\frac{\partial P}{\partial \rho}\right)_\varepsilon - P, \quad \bar\gamma - 1 = \frac{1}{\rho}\left(\frac{\partial P}{\partial \varepsilon}\right)_\rho.$$

Using the binomial approximation for the equation of state considerably simplifies the solution of the problem on discontinuity breakup that underlies our numerical method.

Below, we also give the system of units used in our calculations. The scales of the physical quantities in this system are

$$[r] = R_0, \quad [V_r] = [V_\theta] = [V_\varphi] = (GM_0/R_0)^{1/2}, \tag{10}$$

$$[\rho] = M_0/(4\pi R_0^3), \quad [t] = R_0^{3/2}/(GM_0)^{1/2},$$
$$[P] = GM_0^2/(4\pi R_0^4), \quad [E] = GM_0/R_0,$$
$$[T] = (GM_0^2/(4\pi R_0^4 a_r))^{1/4},$$

where $R_0$ and $M_0$ are some length and mass scales, and $a_r$ is the radiation density constant. In units (10), the hydrodynamic equations with Newtonian gravitation contain no dimensionless parameters. It would be natural to use the following characteristic values from our problem as $R_0$ and $M_0$: $R_0 = 10^8$ cm and $M_0 = 10^{32}$ g. The numerical values for the scales of the physical quantities from (10) are then

$$[r] = 10^8 \text{ cm}, \quad [V_r] = [V_\theta] \tag{11}$$
$$= [V_\varphi] = 2.583 \times 10^8 \text{ cm s}^{-1},$$
$$[\rho] = 7.958 \times 10^6 \text{ g cm}^{-3}, \quad [t] = 3.971 \times 10^{-1} \text{ s},$$
$$[P] = 5.310 \times 10^{23} \text{ erg cm}^{-3}, [E] = 6.674 \times 10^{16} \text{ erg},$$
$$[T] = 2.894 \times 10^9 \text{ K}.$$

*Initial Conditions*

As we noted in the Introduction, the formation time of the PNS embryo with the mass (about $1M_\odot$) characteristic of our problem, when all of the outer layers may be assumed to be in hydrostatic equilibrium, may arbitrarily be taken as the initial time ($t = 0$). We used the distributions of thermodynamic quantities obtained in the studies of the evolution of high-mass stars (Boyes *et al.* 1999) as the initial data for constructing the initial conditions of our problem. From the above paper, we took the density and temperature profiles, which, in turn, serve to determine the initial pressure and internal energy distributions by using the adopted equation of state. Of course, the equation of state used in our calculations differs slightly from the equation of state used in obtaining these profiles. The main difference is that we took into account a large number of nuclides of chemical elements in our numerical calculations of stellar evolution. However, a detailed comparison of the equations of state (Imshennik *et al.* 2002) shows that, quantitatively, these differences are moderately large, and passing to a simpler equation of state in our calculations has no significant effect on the radial profiles of thermodynamic quantities. Nevertheless, the initial profiles were recalculated with the new equation of state. In these calculations, we used the radial density distribution from Boyes *et al.* (1999) and reconstructed the new equatorial distributions of the pressure ($P$) and the remaining thermodynamic quantity $\varepsilon$ by solving the system of hydrostatic equilibrium equations with nonzero rotation (in the form of a centrifugal force):

$$\frac{\partial P}{\partial r} = \rho \left( \frac{V_\varphi^2}{r} - \frac{Gm}{r^2} \right), \tag{12}$$

$$\frac{\partial m}{\partial r} = 4\pi r^2 \rho; \tag{13}$$

the pressure $P$ and the radius $r$ at the inner boundary were also taken from Boyes *et al.* (1999) for $m \approx 1M_\odot$. Thus, strictly speaking, the initial spherically symmetric distribution of matter was one of a hydrostatic equilibrium only in the equatorial plane. The sizes of the computational region were chosen in such a way that the inner boundary was at the radius ($0.876 \times 10^8$ cm) that bounded the region of mass $1M_\odot$ in the initial profile (except the test calculation, in which the inner boundary was a factor of about 2 closer to the center along the radius ($0.401 \times 10^8$ cm)). There was arbitrariness in choosing the outer boundary along the radius. Since the outer, weakly rotating presupernova layers have no fundamental effect on our result, it was not necessary to include them in our analysis. The outer boundary of the computational region in all our calculations was specified at a radius of $10.002 \times 10^8$ cm. The calculations were performed for the evolutionary model of a star with a mass of $25M_\odot$ and solar metallicity ($Z_\odot = 0.02$).

As the initial differential rotation law, we took the analytical formula

$$\omega = \omega_0 \exp\left(-\frac{r^2}{r_0^2}\right), \tag{14}$$

as in our previous paper (Imshennik and Manukovskii 2000), but as a function of the spherical radius. This rotation law corresponds well to the results of the quasi-one-dimensional hydrodynamic calculations of collapse by Imshennik and Nadyozhin (1992), particularly with the given specific values of $\omega_0$ and $r_0$. We emphasize that, according to the data from Boyes *et al.* (1999), the total angular momentum $J_0$ within the iron core is $\sim 10^{50}$ erg s; i.e., it is almost equal to the value from the paper mentioned above. Expression (14) was used to construct the hydrostatic equilibrium distribution of thermodynamic quantities on the equator: $V_\varphi$ in Eq. (12) was calculated by using the formula $V_\varphi = \omega r$. Outside the equator, the initial angular velocity distribution was assumed to be spherically symmetric, in accordance with (14). In contrast, the initial distributions of the azimuthal velocity $V_\varphi$ and the specific angular momentum $j$ were, of course, not spherically symmetric, because they were calculated by using the formulas

$$V_\varphi = \omega r \sin\theta, \quad j = V_\varphi r \sin\theta. \tag{15}$$

In particular, this choice of initial data allows us to avoid such singularities as nonzero azimuthal velocity and angular momentum at $\theta = 0$, and softens a violation of the hydrostatic equilibrium conditions outside the equator.

### *Boundary Conditions*

The region of the solution of our problem, or the computational region, has the shape of a spherical envelope, $r_{\rm min} \le r \le r_{\rm max}$. The choice of $r_{\rm min}$ and $r_{\rm max}$ has already been determined by the physical considerations outlined in the Subsection "Initial Conditions." Sufficient boundary conditions should be specified precisely at these constant boundary values of the Eulerian radius. The inner boundary at $r = r_{\rm min}$ is essentially a transparent wall on which the radial gradients of all physical quantities (**V**, $\rho$, and $\varepsilon$) are assumed to be zero. However, it should be immediately noted that the fact that these derivatives become zero does not adequately fit the physical concept of transparency, being only its simplified, but formally sufficient realization. At the outer boundary, the boundary condition simulates a vacuum outside the computational region: the values of all thermodynamic quantities ($\rho$, $\varepsilon$, and $P$) are set equal to nearly zero at $r = r_{\rm max}$.

Because of equatorial symmetry in addition to axial symmetry, we may restrict our calculations to only one quadrant, so the angle $\theta$ for the computational region ranges from 0 to $\pi/2$. At the $\theta = \pi/2$ boundary, the velocity component $V_\theta$ is assumed to be zero (and again the derivatives of all thermodynamic quantities become zero). The necessary boundary condition for the gravitational potential, $\Phi \to -\dfrac{GM}{r}$ as $r \to \infty$, is automatically satisfied in the algorithm used to solve the Poisson equation (Aksenov 1999).

## RESULTS OF THE NUMERICAL SOLUTION

Below, we describe the results of our numerical solution of the system of differential equations (1)–(5) with the above initial and boundary conditions. In all our post-shock accretion simulations, we used the same model of a high-mass star with a total mass of $25M_\odot$ (Boyes *et al.* 1999). In the main hydrodynamic calculations, the outer computational boundary was exactly at a radius of $1.000168 \times 10^9$ cm, and the inner computational boundary was at a radius of $8.764082 \times 10^7$ cm. The cases differed in the number of zones into which the computational region was divided. In calculation [1], the total number of zones was 100 in the radial direction and 30 in the direction of change of the polar angle; in calculation [2], these numbers were 150 and 45, respectively. The same set of initial data (see the Subsection *Initial Conditions*) was used as the initial presupernova state. To solve the Poisson equation, we expanded the integral representation of the gravitational potential in terms of Legendre polynomials up to the number $l_{\rm max} = 20$.

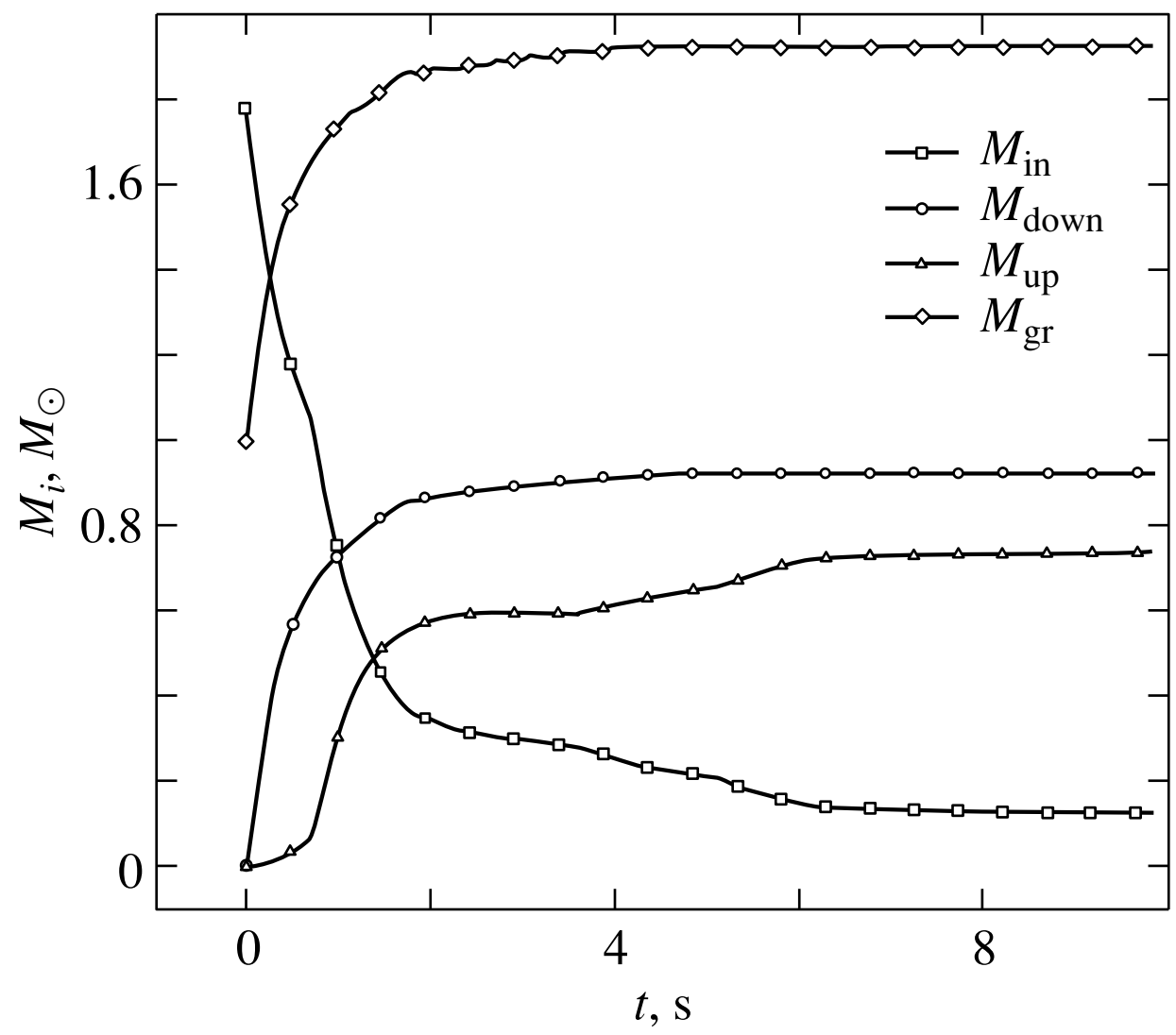

**Fig. 1.** Total mass of the matter versus time ($0 \le \theta \le \pi$): $M_{\rm in}$ is the mass inside the computational region; $M_{\rm down}$ is the mass that passed through the inner boundary $r = r_{\rm min}$; $M_{\rm up}$ is the mass that passed through the outer boundary $r = r_{\rm max}$; $M_{\rm gr}$ is the total mass at $r \le r_{\rm min}$. All data are for calculation [2].

The behavior of the matter was identical in all our calculations. The initially equilibrium (see the Subsection *Initial Conditions*) outer part of the iron core and the outer layers of the presupernova included in the computational region begin to accrete matter. An atmosphere in the form of a torus or, to be more precise, a thick disk is formed near the central PNS during this accretion. As a result of the accretion, part of the matter passes through the transparent inner computational boundary. By the time the toroidal atmosphere may be considered to have been formed, the total mass of the matter that passed inward is the same, with high accuracy, in all our calculations: $\sim 0.93M_\odot$ (Fig. 1). This mass is mainly determined by the choice of the initial rotation law (14) for the matter (see $\omega_0$ and $r_0$ below). Because of the vacuum-simulating artificial boundary condition, the matter at the outer (along the radius) boundary also flows out from the computational region away from the center.

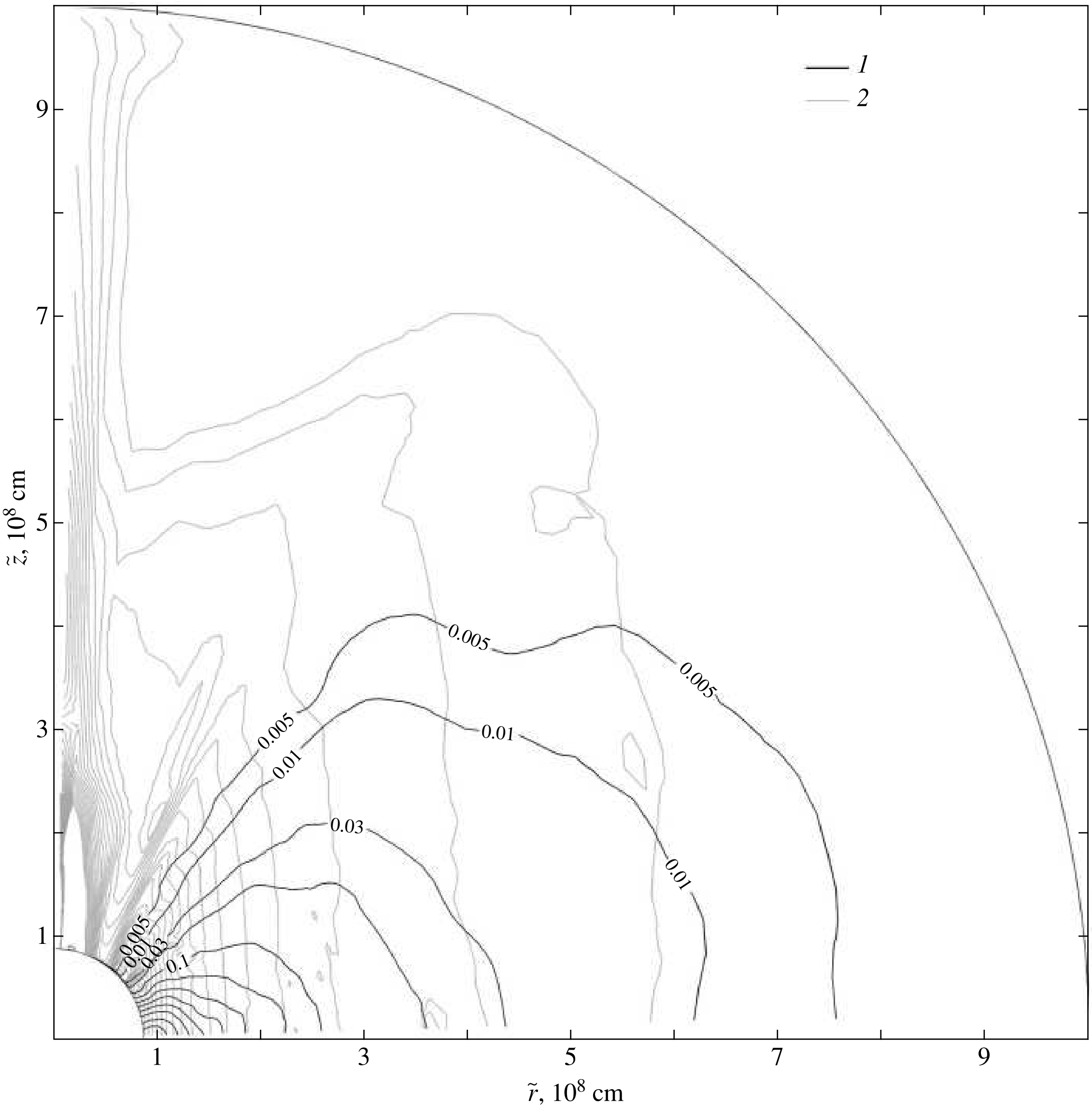

**Fig. 2.** *1*—Lines of equal matter density at the final time for calculation [1] ($t_f = 29.034$ s); *2*—line of equal angular velocity $\omega$ for the same time.

The total mass of the matter that passes outward is $\sim 0.7 M_\odot$ (Fig. 1).

Figures 2 and 3 show lines of equal matter density at the final times for calculations [1] ($t_f = 29.034$ s) and [2] ($t_f = 9.678$ s), respectively. Here (and in Figs. 2–7, 8a, and 9–11), we use the cylindrical coordinates, $\tilde{r} = r \sin\theta$ and $\tilde{z} = r\cos\theta$. Of course, these coordinates are equivalent to the spherical coordinates used in the system of equations (1)–(5) and the finite-difference scheme, but are more convenient for the physical interpretation of the results of our numerical solution. We see that the final state of the matter has an identical structure in both cases. The forming PNS atmosphere is almost twice as extended in the equatorial direction as it is along the rotation axis. For this reason, it would be more precise to call the forming atmosphere a thick disk rather than a torus. The density maximum in both calculations lies in the immediate vicinity of the inner computational boundary. In this situation, the concern about the artificial influence of the chosen boundary condition on the forming atmosphere is quite natural. This circumstance needs the following additional comment.

In an equilibrium rotating atmosphere, the specific angular momentum of the matter is directly proportional to the fluid particle radius. Consequently, particles with a minimum specific angular momentum are located near the inner boundary of the computational region (see below for a discussion of rela-tion (16)). Therefore, if we transfer the inner boundary to a smaller radius, then the formed region will be filled with matter with a lower specific angular momentum; thus, the density maximum will again displace to the inner boundary. The fact that the angular velocity in our initial data varies smoothly ($\omega_0 = 2\ \mathrm{s}^{-1}$, $r_0 = 5 \times 10^8$ cm; see the Subsection *Initial Conditions*), so that there is always an appreciable amount of matter with low specific angular momenta, may also contribute to this displacement. However, the rotation in real high-mass stars is most likely concentrated mainly in the region of the iron core, at the boundary of which the specific angular momentum decreases appreciably. In addition, the concentration of matter near the inner boundary may also be partly attributable to the following factor: In our axisymmetric case $\left(\frac{\partial}{\partial\varphi}, g_\varphi\right) = 0$, the specific angular momentum of each fluid particle must be conserved during its motion. However, the Eulerian scheme used here does not ensure the exact fulfillment of this conservation law. As a result, the specific angular momentum of fluid particles during their predominantly meridional flow can decrease. To check these concerns, we carried out an auxiliary calculation [3].

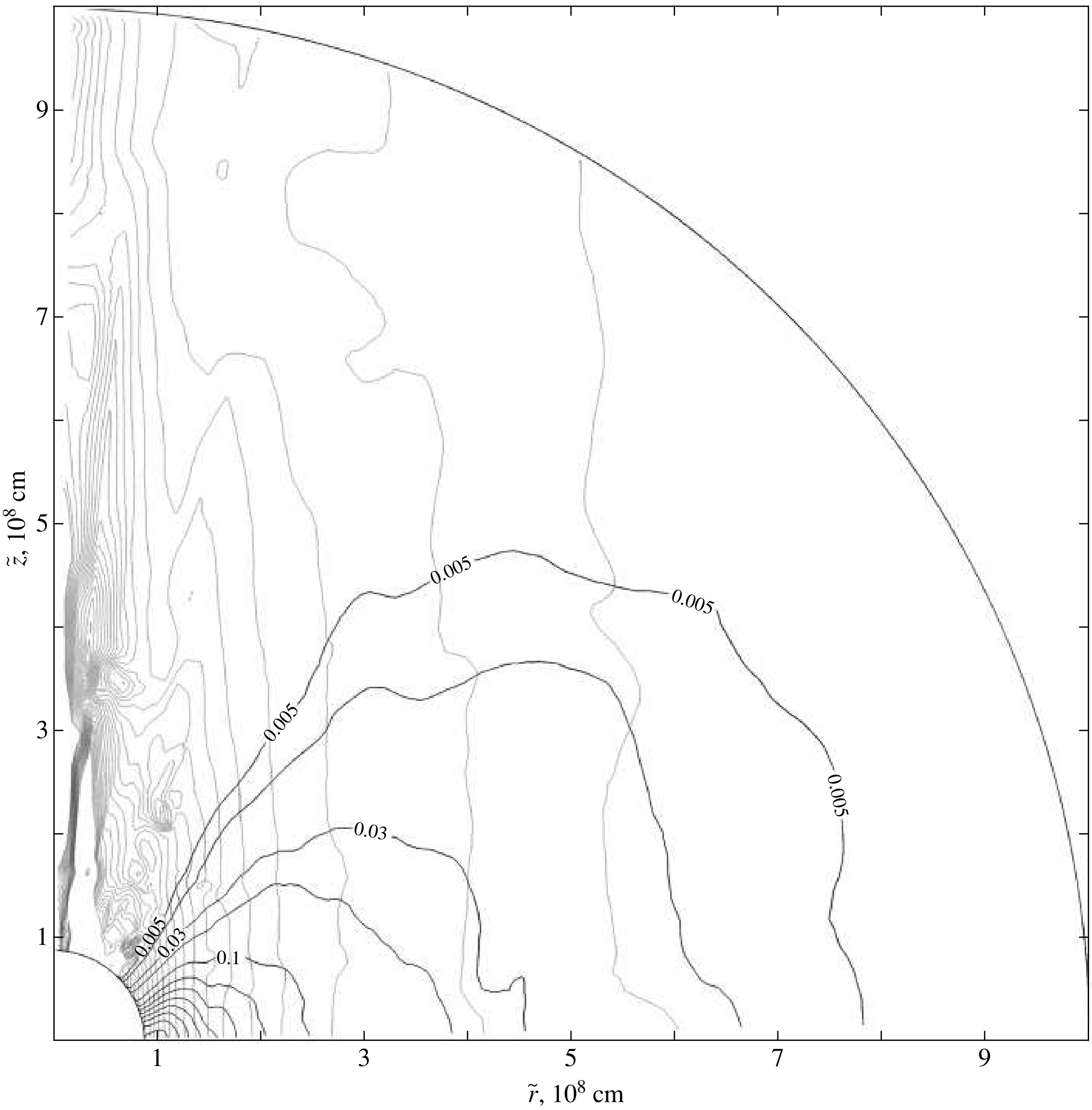

**Fig. 3.** Same as Fig. 2 for calculation [2] ($t_f = 9.678$ s).

In calculation [3], which should be compared with calculation [1], the inner computational boundary was exactly at the radius of $4.010748 \times 10^7$ cm, and the distance to the radius of $8.764082 \times 10^7$ cm was covered by an additional 20 zones (the computational region was divided into 120 cells along the radius). So, in the remaining part of the computational region (from the radius of $8.764082 \times 10^7$ cm to the outer boundary), the division into computational cells coincided exactly with the difference mesh from calculation [1]. The minimum specific angular momentum of the matter from calculation [1] determined at the final time was taken as the minimum admissible value. For matter with a specific angular momentum lower than the value in calculation [3], this value was taken to be identically equal to zero. The data of this calculation are shown in Fig. 4. The lines of equal density, if they are carefully compared with those in Fig. 2, show that the coordinates of the density maximum remained essentially unchanged. As we see from this comparison, the density maximum located in Fig. 2 near a radius $r \approx 0.95$ is clearly established at a radius $r \approx 0.8$ in Fig. 4 for the test calculation. This circumstance removes the above concerns about the decisive influence of the boundary condition specified at the inner boundary on the formation of an atmosphere.

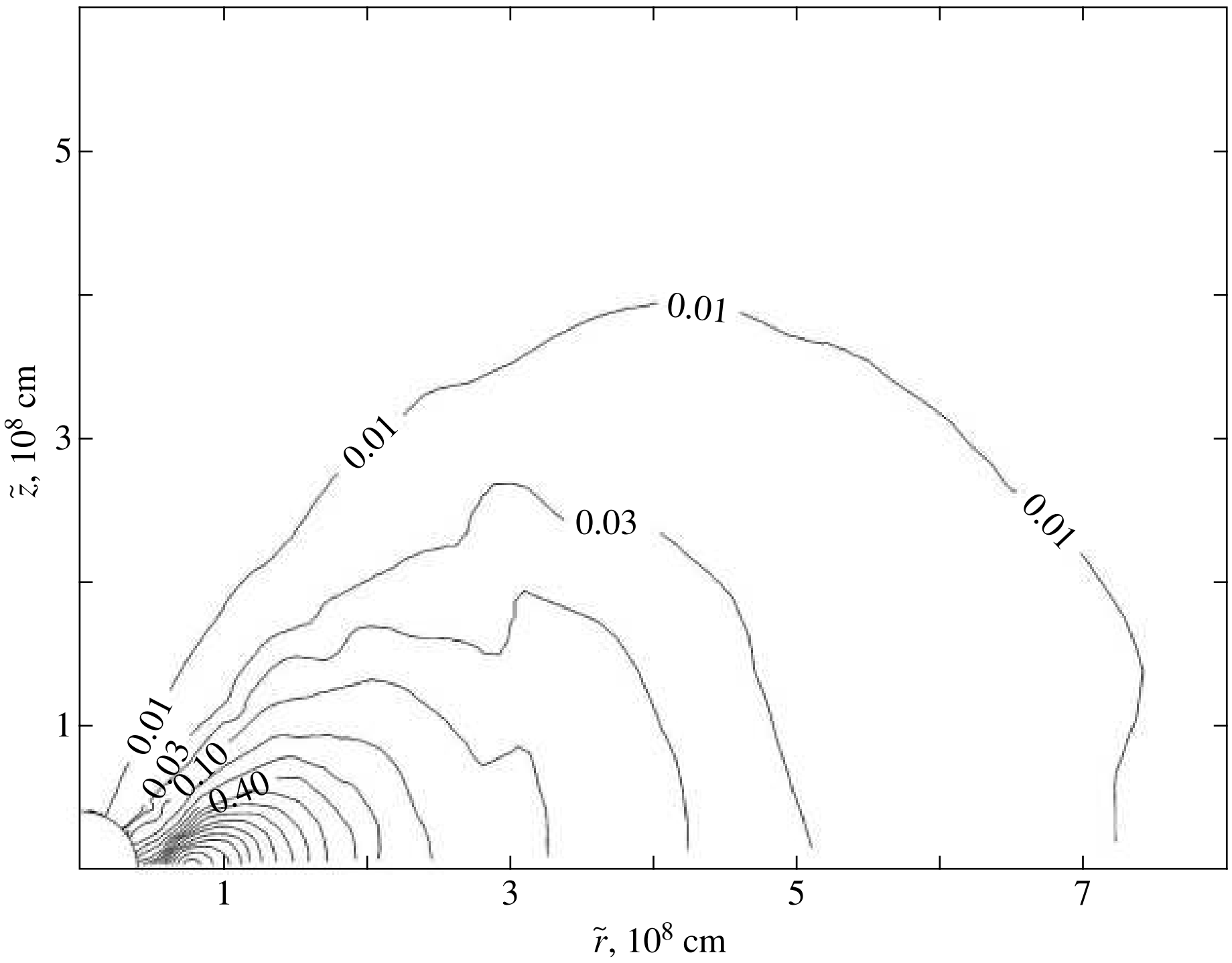

**Fig. 4.** Same as Fig. 2 for the test calculation [3] ($t_f = 6.388$ s). Here, there is no other family of lines of equal angular velocity.

In addition, the results obtained in calculations [1] and [2] should be interpreted as follows. Let $j_{\min}$ be the minimum specific angular momentum of the atmospheric matter in a steady equilibrium state. The simulated steady-state rotating atmosphere is then a result of the post-shock accretion of the outer layers of a high-mass star with the bulk of their rotating matter at the initial time having specific angular momenta higher than the above value of $j_{\min}$. In calculations [1] and [2], the total mass of the matter remaining inside the computational region by the establishment of a steady state is $\sim 0.13 M_\odot$. As we see from the plots in Fig. 1, a steady-state configuration is formed in a time of approximately 6 s; subsequently, the outflow of matter from the computational region virtually ceases, although its weak outflow results from the possible violation of the law of conservation of specific angular momentum. The rates of artificial accretion and ejection due to this circumstance are so low that they have no appreciable effect on the final result of our calculations.

Figure 5 shows lines of constant angular velocity for the final time ($t_f = 9.678$ s) of calculation [2]. As we see from the figure, the angular velocity distribution is cylindrical in pattern; i.e., it depends only on the cylindrical radius $\tilde{r}$ and does not depend on the other cylindrical coordinate $\tilde{z}$. In this case, appreciable deviations from this rotation law appear only in the low-density regions outside the thick disk concentrated near the axis within $\theta \leq \pi/4$. The rotation law $\partial\omega/\partial\tilde{z} = 0$ under discussion and the barotropy of the equation of state $P = P(\rho)$ for steady-state rotating self-gravitating matter are known to be equivalent (Tassoul 1978). The lines of constant (gray) angular velocity were also plotted in Figs. 2 and 3. Their shape also confirms the aforesaid.

More accurate data on the angular velocity distribution for calculation [2] are presented in Fig. 6. In this figure, the angular velocities taken on different rays (their angles are measured from the rotation axis) are plotted against the cylindrical radius $\tilde{r} = r\sin\theta$. As we see, all points satisfactorily fall on the same curve that is best fitted by a power law of the form $\omega = \omega_0^*(r\sin\theta)^\alpha$ rather than an exponential (Gaussian) law of the same argument. Thus, the best fit to the steady-state rotation law in a thick disk is

$$\omega = \omega_0^*(r\sin\theta)^\alpha = 6.0562(r\sin\theta)^{-1.8044}\ \mathrm{s}^{-1}, \quad (16)$$

where the spherical radius $r$ is given in units of the problem $[r] = 10^8$ cm from (11). Note that the exponential (Gaussian) best fit shown in Fig. 6 had $\omega_0 = 13.207\ \mathrm{s}^{-1}$ and $r_0 = 1.706$ (in units of $[r] = 10^8$ cm), but its quality was much lower than that of the power-law fit (16). Thus, we can easily confirm the above assertion that the specific angular momentum is at a minimum near the inner boundary of the computational region. Indeed, $j = \omega r^2\sin^2\theta = \omega_0^*(r\sin\theta)^{0.1956} \propto r^{0.1956}$; i.e., the function increases monotonically but very slowly with radius $r$.

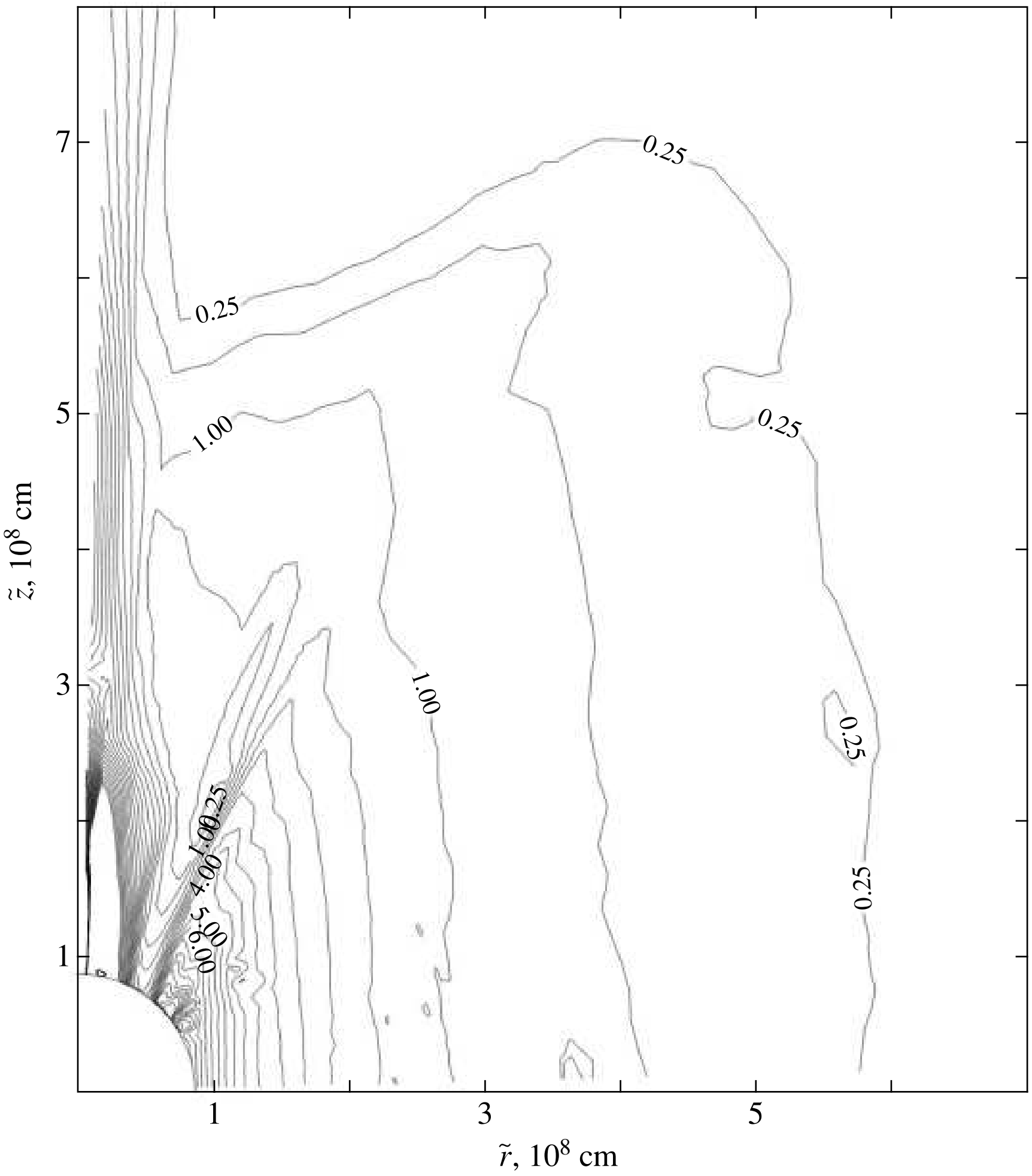

**Fig. 5.** Lines of equal angular velocity $\omega$ at the final time for calculation [2] ($t_f = 9.678$ s).

A remark should also be made about the identification of the presupernova envelope layers that entered into the composition of the toroidal atmosphere rather than collapsing onto the PNS embryo or being ejected outward. The law of conservation of local specific angular momentum allows this identification to be made in principle, because this law is actually violated in the numerical solution (see above). This approach clearly reveals that a rather narrow layer of matter is transferred into the atmosphere from the outer part of the initial iron core and the silicon shell of the presupernova. The assertion that the layer of matter transferred into the toroidal atmosphere is narrow is based on the insignificant change in local angular momentum obtained in our calculation ($j \propto r^{0.2}$). Since the change in atmospheric radius does not exceed a factor of 10, the angular momentum $j$ changes by a factor of $\sim$1.5 (from $j_{\min} \approx 1.51 \times 10^{17}$ to $j_{\max} \approx 2.28 \times 10^{17}\ \mathrm{cm}^2\ \mathrm{s}^{-1}$), while the initial angular momentum changes by a factor much larger than $\sim$10. Figure 7 shows the initial distribution of angular momentum $j$ in accordance with the specified $\omega$ distribution (14) (for the constants $\omega_0$ and $r_0$, see above) in spherical radius $r$. In this figure, the region of matter whose specific angular momentum $j$ lies within the range from $j_{\min}$ to $j_{\max}$ (see above) and which, therefore, could enter the toroidal atmosphere is colored gray. Note that the light areas in Fig. 7 represent the matter that either fell through the inner boundary to the central PNS or was ejected through the outer boundary into outer stellar layers.

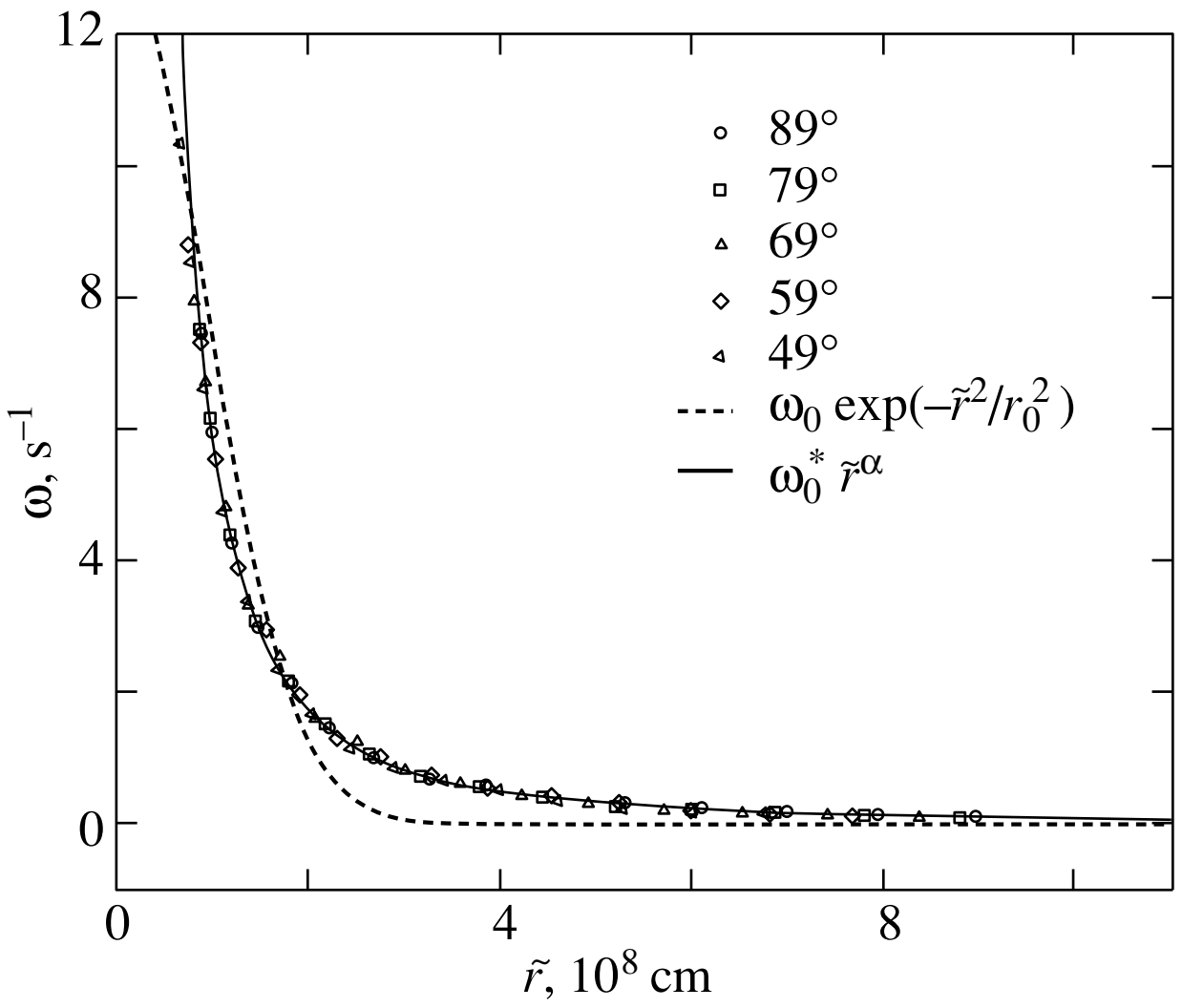

**Fig. 6.** Angular velocity $\omega$ versus cylindrical radius $\tilde{r}$, as constructed from the data on different rays (with a given constant angle $\theta$) at the final time for calculation [2] ($t_f = 9.678$ s): separate points on the rays adjacent to the equator at the location of the thick disk. The analytical fits are represented by the solid (power law) and dashed (exponential) lines. The parameters $\omega_0^*$, $\omega_0$, and $r_0$ are given in the text.

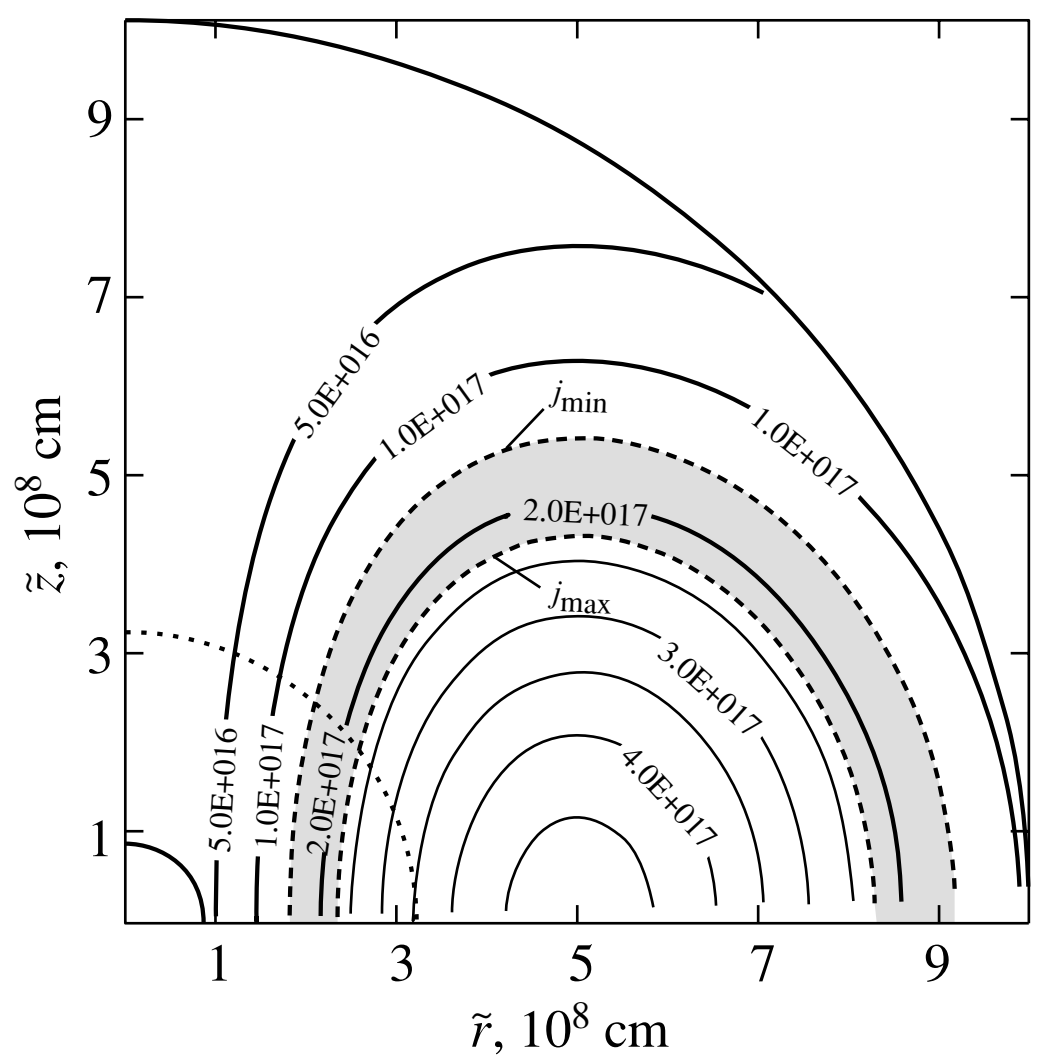

**Fig. 7.** Lines of equal specific angular momentum $j$ at the initial time for calculation [1]. The region of matter whose specific angular momentum lies within the range from $j_{\min} \approx 1.51 \times 10^{17}$ to $j_{\max} \approx 2.28 \times 10^{17}$ cm$^2$ s$^{-1}$ is colored gray; $j_{\min}$ and $j_{\max}$ are the limiting specific angular momenta in the toroidal atmosphere at the final time $t_f =$ 29.034 s). The dotted line represents the boundary of the region of constant entropy.

Let us consider the question of how strongly the derived nonzero temperatures of the matter in a toroidal atmosphere affect the numerical solution. The temperature reaches its maximum near the density maximum of a toroidal atmosphere ($\rho_{\max}$ at $r_{\max}$), $\sim 3 \times 10^9$ K, with solely iron nuclides being present in the nuclear composition of the atmospheric matter. As can be easily verified, this temperature, with the adopted equation of state, leads to a significant temperature correction to the pressure (see Table 1) and specific internal energy of the matter (see Table 2), mainly through an increase in the pressure and specific internal energy of the electron–positron matter and through the appreciable effect of the blackbody radiation at these temperatures. The estimate of the temperature effect by Imshennik and Manukovskii (2000) as $T \sim 2 \times 10^9$ K also qualitatively agrees with its numeric determination in our calculation. Let us discuss the question of whether these temperature corrections are consistent with the requirement that the equation of state for the matter be barotropic. Below, we discuss calculation [1]. Figure 8a shows the initial spherically symmetric distribution of specific entropy $S$ in cylindrical radius $\tilde{r}$ on the equator. The matter that can be in the composition of a toroidal atmosphere (with the corresponding specific angular momentum) is colored gray. The volume in which this matter is contained has the shape of a hollow torus (see Fig. 7). Within this region, the specific entropy takes on two almost constant values: $S_1 \approx 2.4 \times 10^8$ erg g$^{-1}$ K$^{-1}$ up to the radius $r \approx 3.2 \times 10^8$ cm (see Fig. 8a) at which the specific entropy abruptly changes (this radius is marked by the dotted line in Fig. 7) and $S_2 \approx 4.4 \times 10^8$ erg g$^{-1}$ K$^{-1}$ at larger radii. In Fig. 8b, the adiabatic curves with specific entropies $S_1$ and $S_2$ are plotted in the ($\rho$, $T$) plane (the variables $\rho$ and $T$ are on a logarithmic scale). As we see from the figure, in the density range concerned ($10^5$–$2 \times 10^7$ g cm$^{-3}$), the simple Poisson laws with almost constant and similar adiabatic indices $\gamma_1 \approx 1.346$ and $\gamma_2 \approx 1.307$ (see Fig. 8b) hold for $S_1$ and $S_2$, respectively. Thus, particles of matter with two different specific entropies ($S_1$ and $S_2$) could in principle enter a toroidal atmosphere, which would be strange in view of the requirement that the equation of state for the matter of a toroidal atmosphere be barotropic. Interestingly, the total masses of these two regions in Fig. 7 are close

**Table 1.** Pressures of the matter at density $\rho = 3.9 \times 10^6$ g cm$^{-3}$ and two different temperatures ($1.0 \times 10^3$ and $3.3 \times 10^9$ K)

| $T$, K | $P_{\rm rad}$ | $P_{\rm ep}$ | $P_{\rm n}$ | $P_{\rm tot}$ |
|---|---|---|---|---|
| $1.0 \times 10^3$ | $2.500 \times 10^{-3}$ | $1.891 \times 10^{23}$ | $5.791 \times 10^{15}$ | $1.891 \times 10^{23}$ |
| $3.3 \times 10^9$ | $2.904 \times 10^{23}$ | $6.028 \times 10^{23}$ | $1.897 \times 10^{22}$ | $9.122 \times 10^{23}$ |

Note: $P_{\rm rad}$ is the blackbody radiation pressure; $P_{\rm ep}$ is the pressure of the Fermi–Dirac electron–positron gas; $P_{\rm n}$ is the pressure of the perfect Boltzmann gas of nuclides; and $P_{\rm tot}$ is the total pressure of the matter ($P_{\rm tot} = P_{\rm rad} + P_{\rm ep} + P_{\rm n}$). The pressures are given in units of erg cm$^{-3}$.

**Table 2.** Specific internal energies of the matter at density $\rho = 3.9 \times 10^6$ g cm$^{-3}$ and two different temperatures ($1.0 \times 10^3$ and $3.3 \times 10^9$ K)

| $T$, K | $\mathcal{E}_{\rm rad}$ | $\mathcal{E}_{\rm ep}$ | $\mathcal{E}_{\rm n}$ | $\mathcal{E}_{\rm tot}$ |
|---|---|---|---|---|
| $1.0 \times 10^3$ | $1.940 \times 10^{-9}$ | $8.540 \times 10^{16}$ | $2.227 \times 10^{9}$ | $8.540 \times 10^{16}$ |
| $3.3 \times 10^9$ | $2.382 \times 10^{17}$ | $3.800 \times 10^{17}$ | $7.350 \times 10^{15}$ | $6.254 \times 10^{17}$ |

Note: $\mathcal{E}_{\rm rad}$ is the specific blackbody radiation energy; $\mathcal{E}_{\rm ep}$ is the specific energy of the Fermi–Dirac electron–positron gas; $\mathcal{E}_{\rm n}$ is the specific internal energy of the perfect Boltzmann gas of nuclides; and $\mathcal{E}_{\rm tot}$ is the total specific internal energy ($\mathcal{E}_{\rm tot} = \mathcal{E}_{\rm rad} + \mathcal{E}_{\rm ep} + \mathcal{E}_{\rm n}$). The specific internal energies are given in units of erg g$^{-1}$.

**Table 3.** Physical parameters of the numerical solutions for calculations [1–3] of toroidal atmospheres

| Calculation | [1] | [2] | [3] |
|---|---|---|---|
| $t_f$, s | 29.03449 | 9.678182 | 6.387759 |
| $\rho_{\rm max}$, $10^7$ g cm$^{-3}$ | 0.396113 | 0.374559 | 1.152099 |
| $r_{\rm max}$, $10^8$ cm | 0.955023 | 0.981689 | 0.766656 |
| $M^{\rm tot}$, $M_\odot$ | 0.117120 | 0.121641 | 0.211936 |
| $J^{\rm tot}$, $10^{49}$ cm$^2$ s$^{-1}$ g | 4.453701 | 4.716711 | 8.152932 |
| $E^{\rm tot}_{\rm in}$, $10^{50}$ erg | 0.604490 | 0.635241 | 1.203042 |
| $E^{\rm tot}_{\rm gr}$, $10^{50}$ erg | −1.792622 | −1.786542 | −3.893123 |
| $E^{\rm tot}_{\rm kin}$, $10^{50}$ erg | 0.507197 | 0.528444 | 1.228717 |
| $E^{\rm rot}_{\rm kin}$, $10^{50}$ erg | 0.501609 | 0.509417 | 1.169783 |

Note: $t_f$ is the final time of the calculation; $\rho_{\rm max}$ is the density maximum; $r_{\rm max}$ is the cylindrical radius at which the density maximum is located; $M^{\rm tot}$ is the total mass; $J^{\rm tot}$ is the total momentum; $E^{\rm tot}_{\rm in}$ is the total internal energy; $E^{\rm tot}_{\rm gr}$ is the total gravitational energy; $E^{\rm tot}_{\rm kin}$ is the total kinetic energy; $E^{\rm rot}_{\rm kin}$ is the rotational kinetic energy ($\propto V_\varphi^2/2$).

and equal to $M_1 = 0.1136M_\odot$ and $M_2 = 0.1392M_\odot$, respectively. Remarkably, $M_1 \approx M_2 \approx M^{\rm tot}$ (see Table 3) in this case. The data on the specific entropy distribution in the toroidal atmosphere itself, which are plotted in Fig. 9, introduce a larger uncertainty. They convincingly show that only particles with specific entropies in the narrow range $S \approx (3.3–4.5) \times 10^8$ erg g$^{-1}$ K$^{-1}$ entered the toroidal atmosphere. Thus, it is clear that only the second group of particles with the initial specific entropy $S_2$ constituted the toroidal atmosphere obtained in our calculations. At the same time, the first group of particles with the initial specific entropy $S_1$ was accreted onto the PNS embryo (or may have been ejected outward through the outer boundary) in the post-shock accretion process under study. As a result, the equation of state for the matter of a toroidal atmosphere proved to be approximately barotropic, given some errors in the entropy calculation, which, according to Fig. 8b, is quantitatively of little importance.

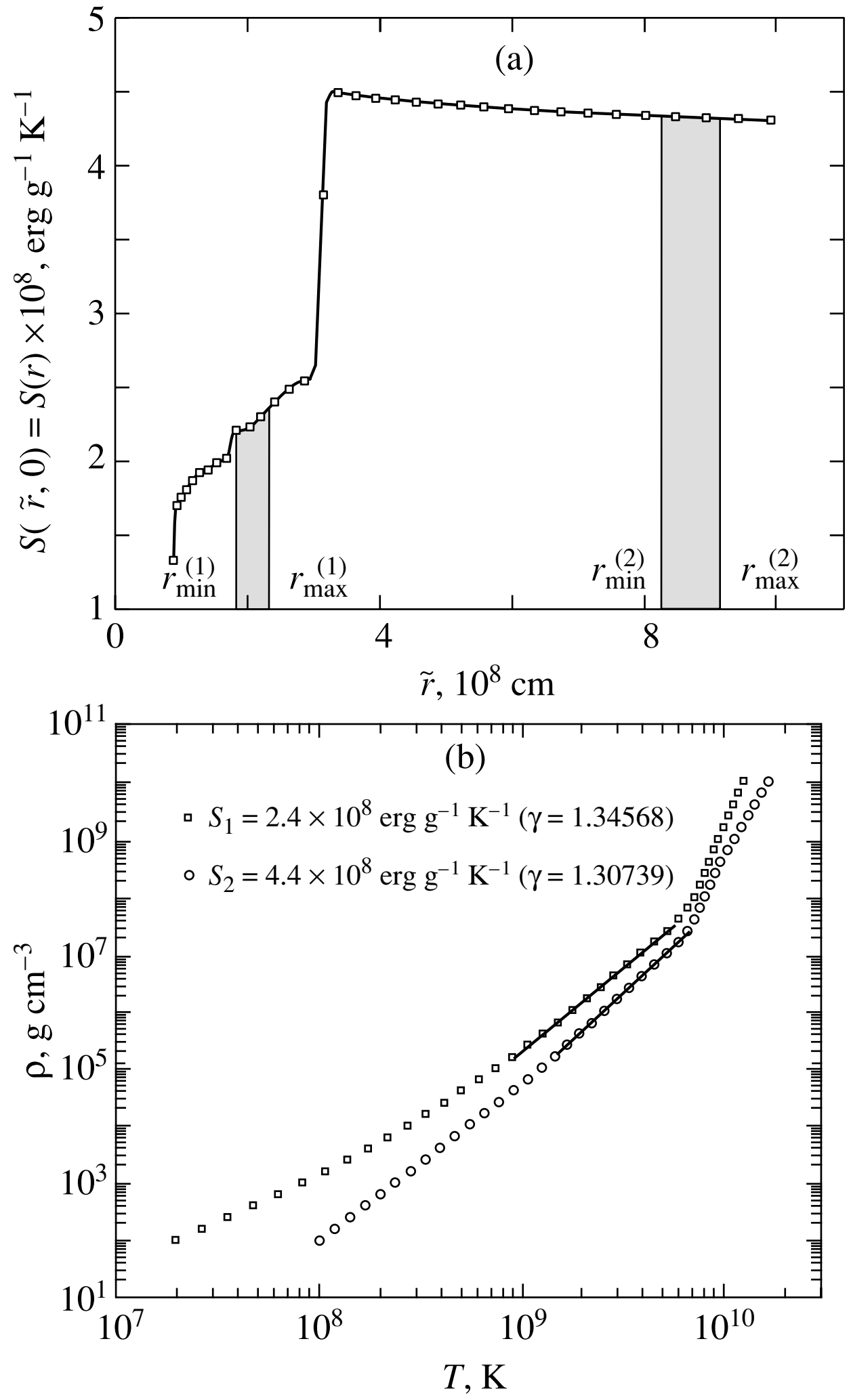

**Fig. 8.** (a) Specific entropy $S$ versus cylindrical radius $\tilde{r}$ on the equator for calculation [1] at the initial time ($r_{min}^{(1)} \approx 1.82 \times 10^8$ cm, $r_{max}^{(1)} \approx 2.35 \times 10^8$ cm and $r_{min}^{(2)} \approx 8.28 \times 10^8$ cm, $r_{max}^{(2)} \approx 9.15 \times 10^8$ cm). (b) The adiabatic curves of the tabulated equation of state for two specific entropies, $S_1 \approx 2.4 \times 10^8$ erg g$^{-1}$ K$^{-1}$ and $S_2 \approx 4.4 \times 10^8$ erg g$^{-1}$ K$^{-1}$. The rectilinear segments are the best linear fits to the adiabatic curves for the density range $10^5$–$2 \times 10^7$ g cm$^{-3}$.

The fact that our numerical solution (see above) is approximately isoentropic (i.e., $\nabla S \cong 0$) (Fig. 9) has a direct bearing on the dynamical stability of toroidal atmospheres. The Viertoft–Lebowitz stability criterion exists for the axisymmetric motions under consideration (Tassoul 1978). In our case (constant entropy), this criterion is

$$\frac{\partial j^2}{\partial \tilde{r}} > 0 \qquad (17)$$

and, as can be easily verified, it is satisfied in accordance with (16). We immediately see from Fig. 9 that the angular velocity distribution in regions with a variable specific entropy ($\nabla S \neq 0$) deviates appreciably from the cylindrically symmetric distribution ($\partial \omega / \partial \tilde{z} \neq 0$). This property is in close agreement with one of the corollaries of the more general formulation of the above criterion (Tassoul 1978).

It is well known that an analytical solution in the form of rotating toroidal atmosphere can be constructed in the axisymmetric case in the Roche approximation for matter with the equation of state of a cold iron gas and arbitrary relativity (Imshennik and Manukovskii 2000). In this case, the dependence on the cylindrical radius alone is the only constraint imposed on the rotation law. If we generalize the analytical solution from the cited paper to the power-law rotation (16), then for our empirically derived rotation law, the formula for the atmospheric density distribution in the plane that passes through the rotation axis is

$$\rho(r,\theta) = B^{-3} \Bigg\{ \Bigg[ \sqrt{1 + B^2 \rho_0^{2/3}} + \frac{GM_0}{M}\left(\frac{1}{r} - \frac{1}{r_0}\right) + \frac{(\omega_0^*)^2}{(2\alpha+2)M} \times \langle (r\sin\theta)^{(2\alpha+2)} - (r_0 \sin\theta_0)^{(2\alpha+2)} \rangle \Bigg]^2 - 1 \Bigg\}^{3/2}, \qquad (18)$$

where $B = 7.792 \times 10^{-3}$ cm g$^{-1/3}$ and $M = 2.272 \times 10^{17}$ cm$^2$ s$^{-2}$ are the constants from the equation of state, $G$ is the gravitational constant, $M_0$ is the mass of the central object, and ($r_0$, $\theta_0$) are the coordinates of the point at which the boundary density $\rho_0$ is specified; the dimension relation for the coefficient $\omega_0^*$ is $[\omega_0] = \lfloor \omega_0^* r_0^{\alpha} \rfloor$. In Fig. 10, the lines of equal density for this analytically specified atmosphere are compared with the lines of equal density for the atmosphere obtained in calculation [1]. The parameter $M_0 = M_{gr} = 1.93 M_{\odot}$ (see Fig. 1) was taken from our calculation. The constants $\rho_0$, $r_0$, and $\theta_0$ were chosen so that the total atmospheric mass from the analytical solution was equal to the numerically calculated atmospheric mass. It should be noted that this choice is ambiguous. However, a common property for any admissible set of these parameters is not only the total mass of the atmosphere but also the position of its density maximum. The latter is determined solely by the properties of the analytical solution and does not depend on these parameters. As we see from Fig. 10, a more compact and denser atmosphere corresponds to the analytical solution. This difference could be explained by the use of different equations of state and by the fact that the meridional flow of matter, which is clearly present in the numerical calculation, was disregarded when deriving the analytical formula. In Fig. 11, atmospheric density is plotted against radius in the equatorial plane. This figure shows the density profiles for both calculations, [1, 2], taken at different times. As we see from the figure, the atmospheric density in calculation [2] (on a finer mesh) is appreciably lower than the density from calculation [1]. The reason is as follows: in calculation [2], at approximately the fifth second of our calculation, we observed the fragmentation of the atmosphere that had already formed by this time; as a result, part of the matter was ejected outward. Concurrently with the fragmentation, the rotation law of the atmosphere changed to a form independent of the cylindrical coordinate $\tilde{z}$. In calculation [1] (on a coarse mesh), we observed no such event; to be more precise, weak perturbations were noticeable near the above time of the calculation (5 s), but they did not eventually lead to its fragmentation. We carried out an additional test numerical calculation [4] on an even finer mesh. The total number of zones in calculation [4] was 200 in the radial direction and 60 in the direction of change of the polar angle. This calculation also confirmed the fragmentation of the atmosphere. In both calculations ([2, 4]), the initial perturbation that arose at the inner boundary of a toroidal atmosphere led to intense ejection of a small fraction of atmospheric matter in the axial direction and to a division of the main volume of the thick disk into two parts approximately equal in mass. As the outer atmospheric fragment receded from the center, part of the matter was transferred back onto the remnant. As a result, the mass of the outward ejected matter did not exceed 30% of the total mass of the thick disk before the onset of fragmentation. The fact that fragmentation in both calculations (calculation [2] and test calculation [4]) began at the same time and developed identically most likely suggests that the observed fragmentation is not numerical in origin.

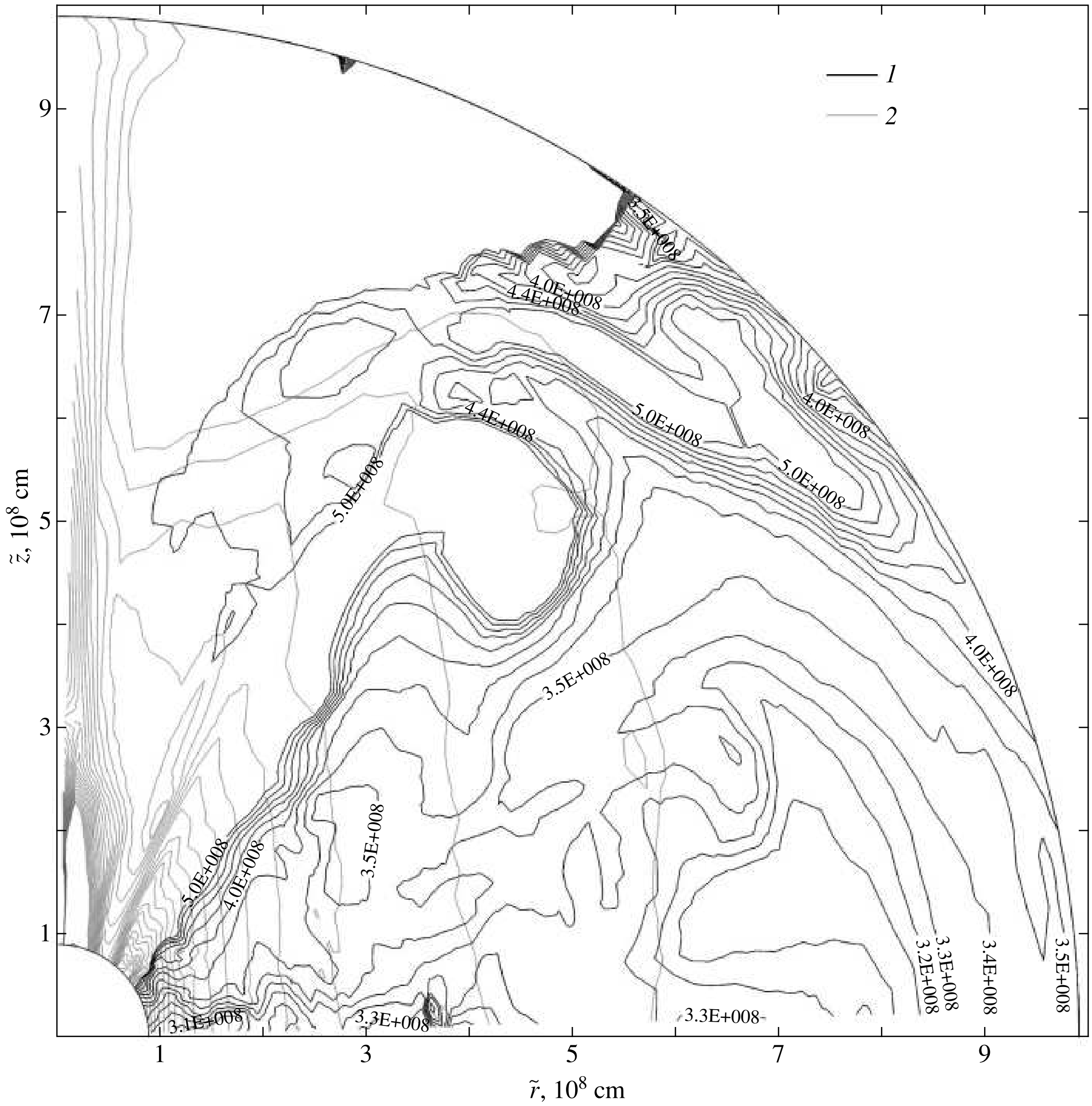

**Fig. 9.** *1*—Lines of equal specific entropy at the final time for calculation [1] ($t_f = 29.034$ s); *2*—lines of equal angular velocity $\omega$ for the same time.

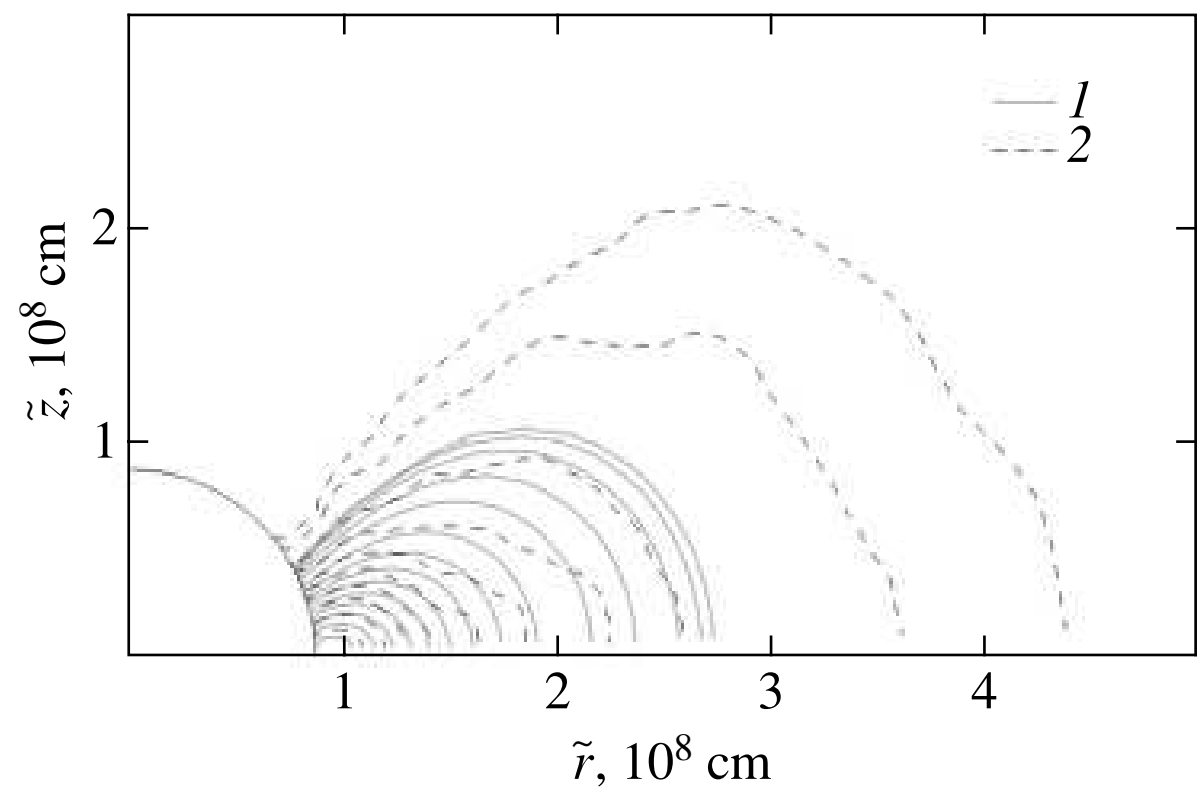

**Fig. 10.** Comparison of the lines of equal atmospheric density at the final time for calculation [1] (*1*) ($t_f =$ 29.034 s) and the lines of equal density for the analytical solution (*2*).

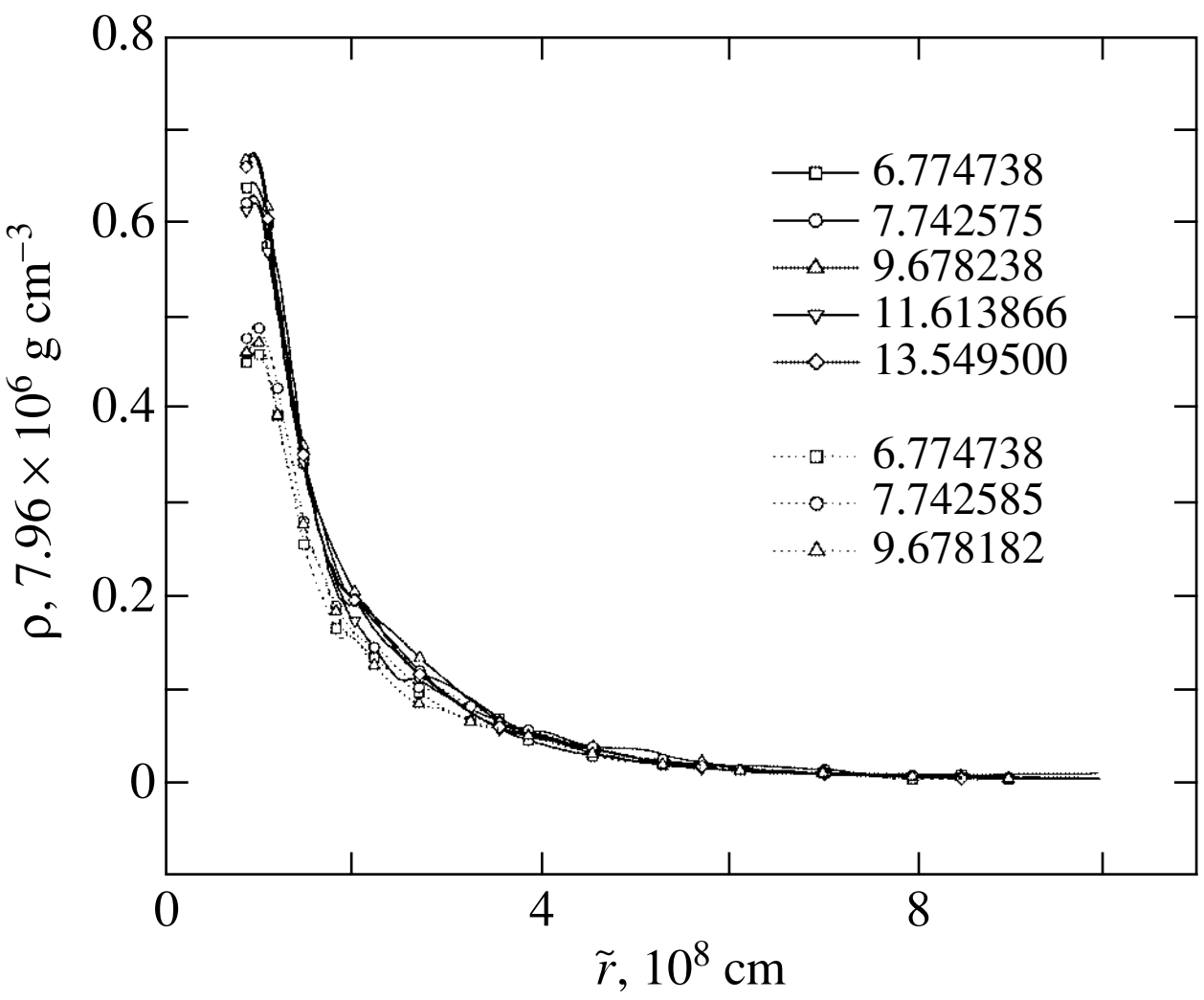

**Fig. 11.** Atmospheric density versus cylindrical radius $\tilde{r}$ in the equatorial plane at various times for calculations [1] (solid lines) and [2] (dotted lines).

Table 3 gives the physical model parameters obtained in calculations [1–3] of toroidal atmospheres. The integrated quantities in the table refer to the entire computational region. Without any particular error, they may be considered to pertain to the atmosphere, because the latter has no well-defined boundary. In addition, its surrounding matter has such a low density that it gives no significant contribution to the integrated quantities. These integrated quantities qualitatively agree with those obtained in the analytical solution (cf. the data from Table 1 for $M = 1.8M_{\odot}$ from Imshennik and Manukovskii (2000)). Of course in this case, there are appreciable differences between the differential rotation laws of the atmospheres being compared (see above in this section). Finally, note that the parameters in calculation [3], in which the inner boundary of the computational region was moved to the center by twice the radius, significantly differ. This change can be mainly explained by the fact that the density maximum is closer to the center and corresponds to a lower specific angular momentum that a substantial mass of the matter has in the initial state; in particular, it can be explained by the shorter computational time. Unfortunately, the influence of the boundary condition itself, which simulates the physical transparency condition for this Eulerian boundary very roughly, cannot be ruled out either. This question probably needs further study.

## CONCLUSIONS

The formation of a toroidal atmosphere near a protoneutron star obtained by the hydrodynamic method may be considered to be the main result of our study. The parameters of this atmosphere with a specified initial rotation law in the outer layers of a high-mass presupernova with a structure from the evolution calculations by Boyes *et al.* (1999) were found to be the following: $M^{\text{tot}} = (0.117\text{–}0.122)M_{\odot}$, $J^{\text{tot}} = (0.445\text{–}0.472) \times 10^{50}$ erg s, $\rho_{\text{max}} = (0.375\text{–}0.396) \times 10^7$ g, and $r_{\text{max}} = (0.955\text{–}0.982) \times 10^8$ cm (calculations [1, 2]). The ranges of these quantities characterize the accuracy of our numerical model for the axisymmetric two-dimensional hydrodynamic problem on post-shock accretion onto the PNS embryo. The atmospheric parameters for the auxiliary calculation [3] differ markedly for the reasons given above (see the Section "Results of the numerical solution"): $M^{\text{tot}} = 0.21M_{\odot}$, $J^{\text{tot}} = 0.82 \times 10^{50}$ erg s, $\rho_{\text{max}} = 1.2 \times 10^7$ g cm$^{-3}$, and $r_{\text{max}} = 0.77 \times 10^8$ cm. Of course, this set of numerical results is attributable to the rotation law specified in the initial conditions. Moreover, this law provided a reasonable PNS mass of $\sim 1.8M_{\odot}$ at the end of our calculation because of the outward ejection of a relatively large mass of $\sim 0.8\text{M}_{\odot}$ and the formation of a toroidal atmosphere with a mass of $\sim 0.2\text{M}_{\odot}$ (the initial mass was equal to the sum of $\sim 1.0\text{M}_{\odot}$ in the PNS embryo and $\sim 1.8\text{M}_{\odot}$ in the outer presupernova layers up to the adopted radius of $\sim 10^9$ cm). This mass distribution between the PNS and the atmosphere was previously obtained in the quasi-one-dimensional calculations of the collapse of an iron stellar core with a mass $M = 2M_{\odot}$ by Imshennik and Nadyozhin (1992). This is not surprising, given the similarity between the rotation laws with the total angular momentum $J_0 \sim 10^{50}$ erg s specified in the initial conditions of the two problems (for the problem within the iron core of mass $M_{\rm Fe} \sim 1.6M_{\odot}$ discussed here). It seems that the chosen rotation law has yet to be justified by gradually developing evolution calculations for rotating high-mass stars until the onset of their collapse, which is undoubtedly a very complex problem. The barotropic properties of atmospheres and the $z$ independence of their angular velocity $\omega$ established in our numerical calculations are sufficient conditions for the validity of the Lichtenstein theorem (see, e.g., Tassoul 1982) that states the existence of an equatorial symmetry plane, a property that is obvious at first glance but far from trivial in reality. Thus, it is quite justifiable to use equatorial symmetry to simplify the numerical solution (see the Subsection *Boundary Conditions*).

It should also be emphasized that the numerically calculated specific parameters of toroidal atmospheres are largely determined by the form of the chosen initial rotation law for the inner presupernova layers (the outer part of the iron core and the silicon shell). In particular, they determine the fact that the matter that was initially in the region of the silicon shell (rather than the iron core) mainly entered the toroidal atmosphere (see the Section "Results of the numerical solution"). In addition, by choosing a different initial rotation law, we could avoid the nonuniformity in the specific entropy distribution of the matter from which a toroidal atmosphere is formed and, as a result, avoid its fragmentation (see above), which may well be associated precisely with this distribution. Nevertheless, the very formation of a toroidal atmosphere during post-shock accretion is a universal phenomenon that depends weakly on this circumstance.

The formulation of the problem on the formation of a toroidal atmosphere was itself largely reinforced by the existence of analytical solutions to the hydrostatic equilibrium equations for a cool iron atmosphere in the gravitational field of a protoneutron star (Imshennik and Manukovskii 2000). Our numerical solution, first, shows the stability of such atmospheres against two-dimensional perturbations (the hydrodynamic sense of the relaxation method!) and, second, removes several restrictions in the formulation of the problem by Imshennik and Manukovskii (2000) due to the allowance for nonzero matter temperature and the effect of self-gravitation, which undoubtedly does not extend beyond minor corrections given the above parameters of a toroidal atmosphere.

## ACKNOWLEDGMENTS

We wish to thank A.V. Zabrodin for his assistance in developing the finite-difference method for solving the problem. This work was supported in part by the Russian Foundation for Basic Research (project no. 00-15-96572) and the Federal Program "Research and Development in Priority Fields of Science and Technology" (contract no. 40.022.1.1.1103).

*Translated by V. Astakhov*